arXiv:1202.0636

# Dust charging processes in the nonequilibrium dusty plasma with nonextensive power-law distribution

Gong Jingyu and Du Jiulin [a]

*Department of Physics, School of Science, Tianjin University, Tianjin 300072, China*

**Abstract** The dust charging processes in the collections of electrons and ions in the nonequilibrium dusty plasma with power-law distributions are investigated on the basic of a new $q$-distribution function theory in nonextensive statistics. Electrons and ions obey the power-law distributions and are with $q$-parameters different from each other. We derive the generalized formulae for the dust charging currents in which the nonextensive effects play roles. Further we investigate the dust charging processes taking place in the homogeneous dusty plasma where only the particle velocities are power-law distributions and in the dust cloud plasma where the particle velocities and densities are both power-law distributions. By numerical analyses, we show that the nonextensive power-law distributions of electrons and ions have significant effects on the dust charging processes in the nonequilibrium dusty plasma.

## I. INTRODUCTION

Dusty plasma contains electrons, ions, and dust grains that are often charged, and they interact by the Coulombian long-range forces. Dusty plasma is a complex system among which the dust grains' charges, masses and sizes are always varying in space and time. The central problem for understanding the nonequilibrium charging processes of dust grains is about the total charging currents, including the electron current and ion current. These currents may be generated by the various ways of contributions coming from the electron collection, the ion collection, the secondary electron emissions, the photoemissions, the thermionic emissions and etc. The dust charging processes have been investigated by theoretical and experimental methods in recent year[1-6]. The related physical phenomena and research works have attracted great interests[7-15].

[a] Electronic mail: jiulindu@yahoo.com.cn

Usually, the dust charging processes are studied theoretically based on the Boltzmann-Gibbs (BG) statistics. However, the dust charging processes does not take place in a system at thermal equilibrium, but usually in a system far away from equilibrium. As we know, spacecraft measurements of plasma velocity distributions, both in the solar wind and in the planetary magnetospheres and magnetosheaths, have revealed that non-Maxwellian distributions are quite common. In many situations the distributions have a "suprathermal" power-law tail at high energies, which has been well modeled by the so called $\kappa$-distribution[16]. It has been reported recently that the plasma system far away from equilibrium may be often with power-law distributions in many physical situations[17, 18], which can be studied in the framework of nonextensive statistics[19], initiated by Tsallis in 1988 as a generalization of BG statistics[0]. The power-law distributions in the nonequilibrium systems with long-range interactions have been studied by the diverse recent and previous statistical theories made for the power-law $q$-distributions[19, 21-25]. The physical explanation for the parameter $q$ different from unity was represented for the nonequilibrium plasma system with Coulombian long-range interactions[19]. Many of the basic characteristics of electron-ion plasmas and dusty plasmas have been investigated under the condition of the power-law $q$-distributions, such as electron and ion dust charging process[26], ion acoustic waves[27-33], dust acoustic waves[34-37], solitary waves[29,36-38], electron acoustic waves[39-41], and Jeans' instability in space plasma[42,43], etc. Nevertheless, according to present knowledge, the usually employed $q$-distribution function is not factorized for kinetic and potential energies because without considering the nonextensivity of the energy. In this case, the $q$-distribution is found to be just an isothermal one, and therefore it might be not a correct description for a nonequilibrium stationary-state of long-range interacting systems[24,25,44]. In order to study the nature of nonequilibrium plasma with power-law distributions, one should seek the new $q$-distribution function.

In this work, using a new form of power-law $q$-distribution function which describe the nonequilibrium stationary state,[24] we investigate the dust charging processes in the collections of electrons and ions in the nonequilibrium dusty plasma where particles are the power-law distributions and obey nonextensive statistical theory. In Sec. II, following the orbital-limited motion (OLM) method[45], we will derive the general formulae of dust charging currents for the nonextensive power-law distributions. In Sec. III, we study the dust charging processes in the homogeneous dust plasma where only the velocities are power-law distributions. And in Sec. IV, we study the dust charging processes in the dust cloud plasma where the velocities and the densities are both power-law distributions. Finally, a conclusion is given in Sec. V.

## II. THE DUST CHARGING CURRENTS

Let us consider finite-sized neutral dust grains immersed in an unmagnetized plasma, and the radius of dust grain is $r_d$. Electrons and ions in the plasma may be absorbed by the dust grains when they approach to the surfaces of dust grains from an infinite distance, where the Coulomb potential of dusty grain plays a role in the interactions when the distance becomes smaller than the Debye radius. The surface potential of the dust grain, $\phi_d$ (a potential relative to average potential of the plasma), is related to its charge, $Q_d$, by $\phi_d = Q_d / r_d$. Thus, the cross section for charging collisions between the dust grains and the ions or electrons can be given[45] as

$$\sigma_j^d = \pi r_d^2 \left(1 - \frac{2Q_j \phi_d}{m_j v_j^2}\right), \tag{1}$$

where the subscript $j=i$, $e$ stands for ion and electron, respectively; $Q_j$ is charge, $v_j$ is velocity and $m_j$ is mass. The cross section $\sigma_j^d$ must be positive so that the collisions can take place, which requires the velocity $v_j \geq \sqrt{2Q_j \phi_d / m_j}$ if $Q_j \phi_d > 0$. As the size of dust grain is usually much smaller than the plasma Debye radius ($r_d \ll \lambda_D$) and it is much smaller than the average inter-grain distance, we can calculate the dust charging current $I_j$, carried by the $j$th component of particles in the plasma. By using the OLM method, the dust charging current can be expressed generally[45] as

$$I_j = Q_j \int_0^{2\pi} d\varphi \int_0^\pi \sin\theta d\theta \int_{v_j^{\min}}^{v_j^{\max}} \sigma_j^d v_j^3 f_j(v_j) dv_j \quad , \qquad (2)$$

where $f_j(v_j)$ is the velocity distribution function of the $j$th component of particles in the plasma. If the velocity distribution function is assumed to be Maxwellian-Boltzmann one, after finishing the integrals in Eq.(2), the dust charging current becomes

$$I_j = 4\pi r_d^2 Q_j n_j \left( \frac{k_B T_j}{2\pi m_j} \right)^{1/2} \left( 1 - \frac{Q_j \phi_d}{k_B T_j} \right) \quad \text{for} \quad Q_j \phi_d < 0 \qquad (3)$$

and

$$I_j = 4\pi r_d^2 Q_j n_j \left( \frac{k_B T_j}{2\pi m_j} \right)^{1/2} \exp\left[ \frac{Q_j \phi_d}{k_B T_j} \right] \quad \text{for} \quad Q_j \phi_d > 0. \qquad (4)$$

It is well known that Maxwellian-Boltzmann distribution only describes the statistical properties of a system being at thermal equilibrium. If a plasma system evolves not to reach a thermal equilibrium due to the Coulombian long-range interactions, but to reach a nonequilibrium stationary state and the particles (electrons, ions or dust grains) have non-Maxwellian-Boltzmann distributions or the power-law distributions[19], the above formulae should need to be modified under a new theoretical framework. We may pay attention to nonextensive statistical mechanics[46], in which the power-law $q$-distribution has been investigated. However, the $q$-distribution function usually used in statistical physics is not factorized for kinetic and potential energies because without considering the nonextensivity of energy. In the light of present understanding,[24,25,44] this old $q$-distribution function is just an isothermal distribution and therefore it might not describe a nonequilibrium stationary state of long-range interacting systems. Here we will employ a new $q$-distribution function that is factorized for kinetic and potential energies, where the nonextensivity of energy has been taken into consideration. This new nonextensive power-law $q$-distribution is known to describe some of statistical properties of a long-range interacting system being at nonequilibrium stationary-state[19,21,23-25]. The power-law $q$-distribution function for a nonequilibrium dynamical many-body system[24,25] is given by

$$f_j(v_j, r_j) = A_{qj}\, n_j(r_j)\left[1 - (1 - q_j)\frac{m_j v_j^2}{2k_B T_j}\right]^{1/(1-q_j)}.$$  (5)

This distribution function could excellently model the dark matter haloes observed on spherical galaxies with regard to their nonequilibrium stationary states[47], and it has been applied to construct the reaction rate theory for the nonequilibrium systems with power-law distributions.[48] In Eq.(5), there is a truncation for the velocity if the nonextensive parameter is $q_j < 1$, i.e. $v_j \leq \sqrt{2k_B T_j / (1 - q_j) m_j}$, but there is no truncation if it is $q_j > 1$. Here for the plasma with Coulombian long-range interactions we can express the number density of a component as

$$n_j(r_j) = n_{j0}\left[1 - (1 - q_j)\frac{Q_j \phi(r_j)}{k_B T_j}\right]^{\frac{1}{1-q_j}},$$  (6)

where $q_j$ is the parameter of $j$th component, and $\phi(r_j)$ is the electrostatic field potential; $A_{qj}$ is the normalization constant given by

$$A_{qj} = \begin{cases} \dfrac{(5 - 3q_j)(3 - q_j)\sqrt{1 - q_j}}{4}\left[\dfrac{m_j}{2\pi k_B T_j}\right]^{\frac{3}{2}} \cdot \dfrac{\Gamma\left(\dfrac{1}{1-q_j} + \dfrac{1}{2}\right)}{\Gamma\left(\dfrac{1}{1-q_j}\right)}, \text{ for } q_j < 1 \\[6mm] \dfrac{(5 - 3q_j)\sqrt{q_j - 1}}{2}\left[\dfrac{m_j}{2\pi k_B T_j}\right]^{\frac{3}{2}} \cdot \dfrac{\Gamma\left(\dfrac{1}{q_j - 1}\right)}{\Gamma\left(\dfrac{1}{q_j - 1} - \dfrac{1}{2}\right)}, \text{ for } q_j > 1 \end{cases},$$  (7)

Substituting the distribution function Eq.(5) and the cross section Eq.(1) into Eq.(2), the dust charging current becomes

$$I_j = 4\pi^2 Q_j r_d^2 n_j(r_j) A_{qj} \int_{v_j^{\min}}^{v_j^{\max}} dv_j v_j^3 \left(1 - \frac{2Q_j \phi_d}{m_j v_j^2}\right)\left[1 - (1 - q_j)\frac{m_j v_j^2}{2k_B T_j}\right]^{\frac{1}{(1-q_j)}}.$$  (6)

The integral bounds for the velocity can be determined easily when the restrictions required for the velocity in Eq.(1) and Eq.(5) are taken into consideration. For the sake of economization of the space, here we may use $F_j(v_j, q_j)$ to denote the integrand in Eq.(8). Thus if $Q_j \phi_d < 0$ we have

$$I_j = 4\pi^2 Q_j r_d^2 n_j(r_j) A_{qj} \int_0^{\sqrt{2k_B T_j/(1-q_j)m_j}} dv_j F_j(v_j, q_j) \quad \text{for } q_j < 1, \tag{9}$$

and

$$I_j = 4\pi^2 Q_j r_d^2 n_j(r_j) A_{qj} \int_0^\infty dv_j F_j(v_j, q_j) \quad \text{for } q_j > 1. \tag{10}$$

And if $Q_j \phi_d > 0$ we have

$$I_j = 4\pi^2 Q_j r_d^2 n_j(r_j) A_{qj} \int_{\sqrt{2Q_j\phi_d/m_j}}^{\sqrt{2k_B T_j/(1-q_j)m_j}} dv_j F_j(v_j, q_j) \quad \text{for } q_j < 1, \tag{11}$$

and

$$I_j = 4\pi^2 Q_j r_d^2 n_j(r_j) A_{qj} \int_{\sqrt{2Q_j\phi_d/m_j}}^\infty dv_j F_j(v_j, q_j) \quad \text{for } q_j > 1. \tag{12}$$

After finishing the integrals in Eqs.(9)-(12), the dust charging current becomes

$$I_j = \sqrt{8\pi} r_d^2 n_j(r_j) Q_j B_{qj} \sqrt{\frac{k_B T_j}{m_j}} \left[ 1 - (3 - 2q_j) \frac{Q_j \phi_d}{k_B T_j} \right] \quad \text{for } Q_j \phi_d < 0, \tag{13}$$

and

$$I_j = \sqrt{8\pi} r_d^2 n_j(r_j) Q_j B_{qj} \sqrt{\frac{k_B T_j}{m_j}} \left[ 1 - (1 - q_j) \frac{Q_j \phi_d}{k_B T_j} \right]^{(3-2q_j)/(1-q_j)} \quad \text{for } Q_j \phi_d > 0, \tag{14}$$

where the parameter $B_{qj}$ is given by

$$B_{qj} = \begin{cases} \dfrac{(5-3q_j)(3-q_j)\sqrt{1-q_j}}{4(2-q_j)(3-2q_j)} \cdot \dfrac{\Gamma\left(\dfrac{1}{1-q_j}+\dfrac{1}{2}\right)}{\Gamma\left(\dfrac{1}{1-q_j}\right)}, \text{for } q_j < 1 \\ \\ \dfrac{(5-3q_j)\sqrt{q_j-1}}{2(2-q_j)(3-2q_j)} \cdot \dfrac{\Gamma\left(\dfrac{1}{q_j-1}\right)}{\Gamma\left(\dfrac{1}{q_j-1}-\dfrac{1}{2}\right)}, \text{ for } q_j < 1 \end{cases} \tag{15}$$

Generally speaking, the electron current is much bigger than the ion current on account of mass of the ion is much larger than mass of the electron, and then the dust grains are negatively charged. In this case, the dust grain charge, the ion charge and the electron charge are $Q_d = -Z_d e$ ( $Z_d > 0$ ), $Q_i = e$ and $Q_e = -e$ , respectively. Then we get the dust charging currents for the ions,

$$I_i = \sqrt{8\pi} r_d^2 n_i(r_i) e B_{qi} \sqrt{\frac{k_B T_i}{m_i}} \left[ 1 - (3 - 2q_i) \frac{e\phi_d}{k_B T_i} \right],  \qquad (16)$$

and for the electrons,

$$I_e = -\sqrt{8\pi} r_d^2 n_e(r_e) e B_{qe} \sqrt{\frac{k_B T_e}{m_e}} \left[ 1 + (1 - q_e) \frac{e\phi_d}{k_B T_e} \right]^{(3 - 2q_e)/(1 - q_e)},  \qquad (17)$$

## III. DUST GRAIN CHARGE IN HOMOGENEOUS PLASMA

We first consider homogeneous dusty plasma. In Eq.(6), if there is $Q_j \phi(r_j) << k_B T_j$ , the densities of ions and electrons, $n_e$ and $n_i$ , are constants approximately. The dust grains are negatively charged. With the negative potential increase at the surface of dust grains, the ion current approaching to the surface increases gradually but the electron current decreases, finally satisfying the current equilibrium: $I_i + I_e = 0$ . According to this equilibrium as well as the charge quasi-neutral condition, i.e. $n_e + Z_d n_d = n_i$ , from Eqs.(16) and (17）we find

$$\frac{B_{qi}}{B_{qe}} \sqrt{\frac{T_i m_e}{T_e m_i}} \left[ 1 - (3 - 2q_i) \frac{e\phi_d}{k_B T_i} \right] \left[ 1 + (1 - q_e) \frac{e\phi_d}{k_B T_e} \right]^{-(3 - 2q_e)/(1 - q_e)} = 1 - Z_d \frac{n_d}{n_i}.  \qquad (18)$$

On the right hand side of this equation, $Z_d$ can be replaced by $\phi_d = -Z_d e/r_d$ , and we let $T_i/T_e = \sigma$ . Then the equation is written as

$$\left[ 1 - (3 - 2q_i) U \right] - \frac{B_{qe}}{B_{qi}} (1 + PU) \sqrt{\frac{m_i}{\sigma m_e}} \left[ 1 + (1 - q_e) \sigma U \right]^{(3 - 2q_e)/(1 - q_e)} = 0,  \qquad (19)$$

where we have denoted $U = e\phi_d/k_B T_i$ , called as normalized dust grain surface potential, and $P = 4\pi n_d r_d \lambda_{D0}^2$ with a parameter $\lambda_{D0} = \sqrt{k_B T_i/4\pi n_i e^2}$ , called as normalized particle number density.

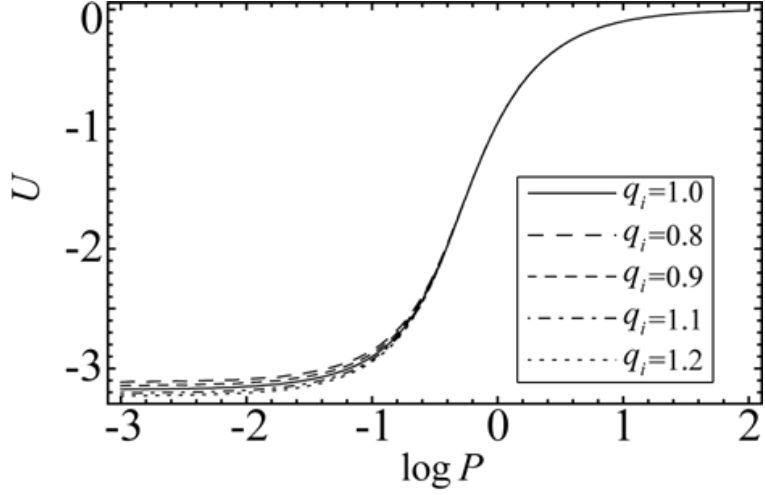

Fig.1. The relation between $U$ and $P$ for different values of $q_i$ under the condition of $m_i/m_e = 10^4$, $\sigma = 1$ and $q_e = 1$.

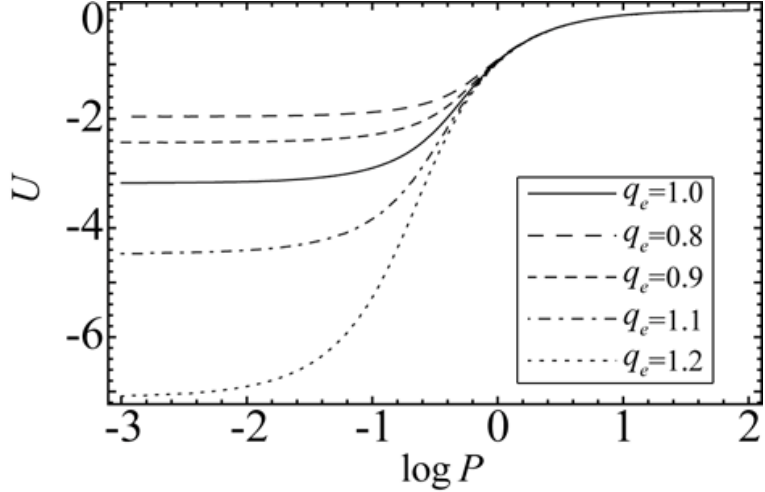

Fig.2. The relation between $U$ and $P$ for different values of $q_e$ under the condition of $m_i/m_e = 10^4$, $\sigma = 1$ and $q_i = 1$.

Eq.(19) determines a relation that depends on $q_j (j = i, e)$ between $U$ and $P$ in the dusty plasma with the nonextensive power-law distributions, which is equivalent to the relation between the number density of dust grains and the surface potential of dust grain, $\phi_d$. By performing numerical calculations, this relation can be shown in Fig.1 and Fig.2 for different values of the nonextensive parameters $q_i$ and $q_e$. We find that the nonextensivity has almost no any effect on the relation between the normalized surface potential $U$ and the normalized density $P$ when $\log P$ is more than zero, but it has significant effect on this relation between $U$ and $P$ only when log $P$ is less than zero. In particular, this relation depends intensively on the nonextensive parameter of electrons, $q_e$, different from unity when log $P$ is less than zero.

## IV. DUST GRAIN CHARGE IN INHOMOGENEOUS DUST CLOUD

Following the dust cloud model of Havnes et al[49], we consider a finite-sized dust cloud embedded in an infinite-sized plasma. The potential of the dust cloud is different from that of the ambient plasma. The particle sources of plasma are at a distance infinite far from the dust cloud. The temperatures of electrons and ions are assumed to be not affected by the dust cloud potential. The dust cloud potential is denoted by $\phi_c$ (the electrostatic potential), and the floating potential of the dust grains in the cloud is denoted by $\phi_f = -Z_d e / r_d$. The floating potential refers to potential at the surface of gust grain where no net current (the sum of negative and positive current is zero) flows. In other word, if a dust grain is immersed in electron-ion plasma and the ion current and the electron current at the grains' surface attain a balance, then the dust grain has the floating potential. We still consider the dust grains in the case of negatively charged. The dust grain charge, the ion charge and the electron charge are $Q_d = -Z_d e$ ($Z_d > 0$), $Q_i = e$ and $Q_e = -e$, respectively. When the dust charging currents get a balance in the dust cloud, i.e. satisfying the equilibrium condition, $I_i + I_e = 0$, the surface potential $\phi_d$ in Eqs.(16) and (17) should be replaced by the floating potential $\phi_f$. Namely,

$$I_i = \sqrt{8\pi} r_d^2 n_i(r_i) e B_{qi} \sqrt{\frac{k_B T_i}{m_i}} \left[ 1 - (3 - 2q_i) \frac{e\phi_f}{k_B T_i} \right], \qquad (20)$$

$$I_e = -\sqrt{8\pi} r_d^2 n_e(r_e) e B_{qe} \sqrt{\frac{k_B T_e}{m_e}} \left[ 1 + (1 - q_e) \frac{e\phi_f}{k_B T_e} \right]^{(3-2q_e)/(1-q_e)}. \qquad (21)$$

Ions and electrons in the cloud are now inhomogeneous and their densities obey the nonextensive power-law distribution Eq.(6), given by

$$n_i(r_i) = n_{i0} \left[ 1 - (1 - q_i) \frac{e\phi_c(r_i)}{k_B T_i} \right]^{\frac{1}{1-q_i}},$$ (22)

$$n_e(r_e) = n_{e0} \left[ 1 + (1 - q_e) \frac{e\phi_c(r_e)}{k_B T_e} \right]^{\frac{1}{1-q_e}},$$ (23)

where $n_{i0} = n_{e0} = n_0$ stands for the density of ions and electrons at infinite distance.

The charge quasi-neutral condition, $n_e + Z_d n_d = n_i$, becomes

$$\left[ 1 + (1 - q_e) \frac{e\phi_c}{k_B T_e} \right]^{\frac{1}{1-q_e}} - \left[ 1 - (1 - q_i) \frac{e\phi_c}{k_B T_i} \right]^{\frac{1}{1-q_i}} - 4\pi r_d \lambda_{Di}^2 n_d \left( \frac{e\phi_f}{k_B T_i} \right) = 0$$ (24)

with $\lambda_{Di} = \sqrt{k_B T_i / 4\pi n_0 e^2}$. By using the equilibrium condition, $I_i + I_e = 0$, we derive

that

$$\frac{B_{qi}}{B_{qe}} \sqrt{\frac{T_i m_e}{T_e m_i}} \frac{\left[ 1 - (3 - 2q_i) \frac{e\phi_f}{k_B T_i} \right]}{\left[ 1 + (1 - q_e) \frac{e\phi_f}{k_B T_e} \right]^{(3-2q_e)/(1-q_e)}} = \frac{\left[ 1 + (1 - q_e) \frac{e\phi_c}{k_B T_e} \right]^{1/(1-q_e)}}{\left[ 1 - (1 - q_i) \frac{e\phi_c}{k_B T_i} \right]^{1/(1-q_i)}}.$$ (25)

The potentials $\phi_c$ and $\phi_f$ can be obtained by solving Eqs.(24) and (25). Now we let

$\sigma = T_i / T_e$, $P = 4\pi n_d r_d \lambda_{Di}^2$ ( the normalized density), $U_c = e\phi_c / k_B T_i$ ( the normalized

dust cloud potential), and $U_f = e\phi_f / k_B T_i$ ( the normalized dust floating potential),

Eqs.(24) and (25) can be written as

$$\left[ 1 + (1 - q_e)\sigma U_c \right]^{\frac{1}{1-q_e}} - \left[ 1 - (1 - q_i) U_c \right]^{\frac{1}{1-q_i}} - P U_f = 0,$$ (26)

and

$$\frac{B_{qi}}{B_{qe}} \sqrt{\frac{\sigma m_e}{m_i}} \frac{\left[ 1 - (3 - 2q_i) U_f \right]}{\left[ 1 + (1 - q_e)\sigma U_f \right]^{(3-2q_e)/(1-q_e)}} = \frac{\left[ 1 + (1 - q_e)\sigma U_c \right]^{1/(1-q_e)}}{\left[ 1 - (1 - q_i) U_c \right]^{1/(1-q_i)}}.$$ (27)

On the basis of Eqs.(26) and (27), we can determine the relation between the dust cloud potential $U_c$ and the density $P$ as well as the relation between the dust floating potential $U_f$ and the density $P$, which are equivalent to the relations between the density $n_d$ and the potentials $\phi_c$ and $\phi_f$. These relations now depend on the nonextensive parameters $q_i$ and $q_e$. By performing numerical calculations, these relations can be illustrated in Figs.3-6.

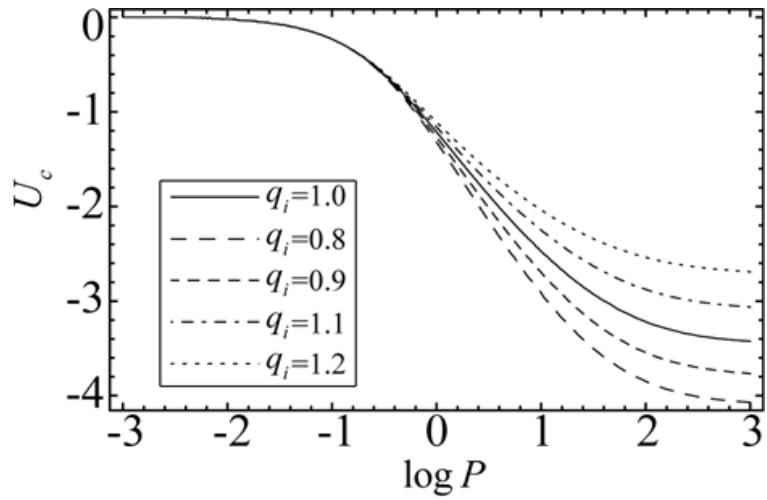

Fig.3. The relation between $U_c$ and $P$ for different values of $q_i$ under the

Condition of $m_i/m_e = 10^4$, $\sigma = 1$ and $q_e = 1$.

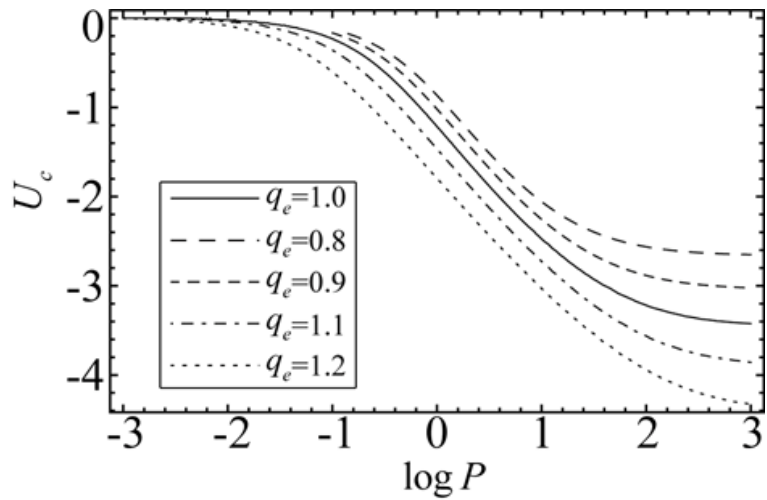

Fig.4. The relation between $U_c$ and $P$ for different values of $q_e$ under the condition

of $m_i/m_e = 10^4$, $\sigma = 1$ and $q_i = 1$.

Fig.3 and Fig.4 show the relations between $U_c$ and $\log P$ for different values of $q_i$ and $q_e$ , respectively, if we take $\sigma = 1$ and $m_i/m_e = 10^4$ . It is found that the nonextensive parameter $q_i$ has a significant effect on the relation between $U_c$ and $\log P$ only when $\log P$ is more than zero, and this effect becomes more and more obvious with the increase of $\log P$ , while the effect of nonextensive parameter $q_e$ on the relation grows gradually with the increase of $\log P$ .

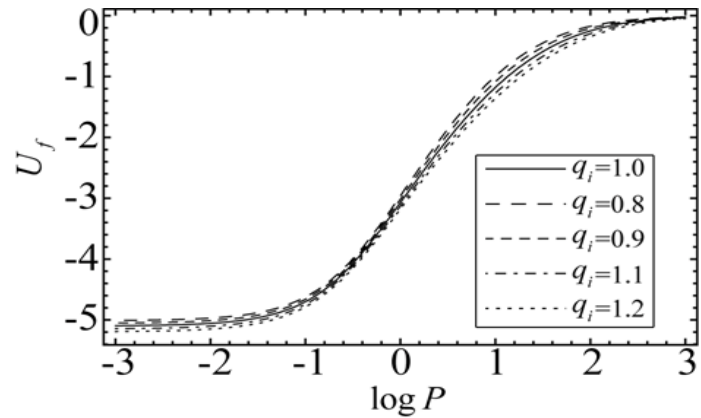

Fig.5. The relation between $U_f$ and $P$ for different values of $q_i$ under the condition of $m_i/m_e = 10^4$ , $\sigma = 1$ and $q_e = 1$ .

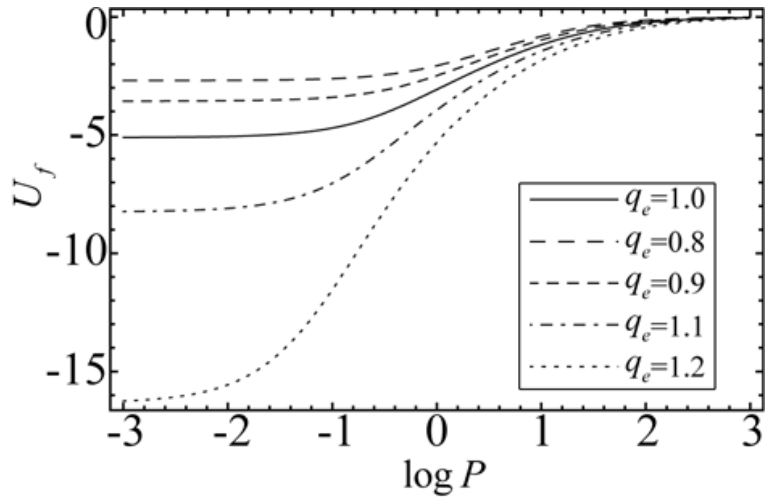

Fig.6. The relation between $U_f$ and $P$ for different values of $q_e$ under the condition of $m_i/m_e = 10^4$ , $\sigma = 1$ and $q_i = 1$ .

Fig.5 and Fig.6 show the relation between $U_f$ and $\log P$ for different values of $q_i$ and $q_e$ , respectively, if we take $\sigma = 1$ and $m_i/m_e = 10^4$ . We find that the nonextensive parameter $q_i$ has a small effect generally on the relation between $U_f$ and $\log P$ , while the nonextensive parameter $q_e$ has a significant effect on this relation when $\log P$ is less than zero, and this effect becomes more and more obvious with the decrease of $\log P$ .

## V. CONCLUSION

In summary, we have studied the dust charging process via the collections of electrons and ions in the dusty plasma where particles are assumed to be the power-law distributions and to obey a new nonextensive statistical theory (see Eqs.(5) and (6)). In particular, we have employed a new form of $q$-distribution function that is factorized for kinetic and potential energies due to the nonextensivity of the energy in the dusty plasma. This new nonextensive power-law $q$-distribution is known to describe the nonequilibrium stationary state of a system away from equilibrium. The electrons and the ions have the nonextensive parameters, $q_i$ and $q_e$, different from each other and then they have different effects on the dust charging process. Firstly, by using the usual OLM approach, we can get new formulae for the dust charging currents in the dusty plasma with the nonextensive power-law q-distributions (see Eqs.(13) and (14)). Secondly, we investigated the dust charging process in the homogeneous dusty plasma with the nonextensive power-law q-distributions, where the densities of electrons and ions are constants approximately. Thirdly, we investigated the dust charging process in the dust cloud model (Havnes $et$ $al$[49],1987) with the nonextensive power-law q-distributions, where the velocities and densities of the particles in the plasma are both the power-law q-distributions. For the dust charging process in the homogeneous dusty plasma, we have derived a relation between the normalized dust grain surface potential $U$ and the normalized particle number density $P$ (see Eq.(19)), which depends on the nonextensive parameters of the electrons and ions, $q_j (j = i, e)$. For the dust charging process in the dust cloud model, we have also derived a relation between the dust cloud potential $U_c$ and the density $P$ as well as a relation between the dust floating potential $U_f$ and the density $P$ (see Eqs.(26) and (27)), which also depends on the nonextensive parameters of the electrons and ions, $q_j (j = i, e)$.

By performing numerical analyses, we have showed that the relation between $U$ and $P$ for the homogeneous dusty plasma depends intensively on the nonextensive parameter of electrons, $q_e$, when $\log P$ is less than zero. For the dust cloud model, it has been shown that the nonextensive parameter of ions, $q_i$, has a significant effect on the relation between $U_c$ and $P$ only when $\log P$ is more than zero, and this effect becomes more and more obvious with the increase of $\log P$, while the effect of the nonextensive parameter of electrons, $q_e$, on the relation grows generally with the increase of $\log P$. The nonextensive parameter of electrons, $q_e$, has a significant effect on the relation between $U_f$ and $P$ and this effect becomes more and more obvious with the decrease of $\log P$. In conclusion, the nonextensive power-law distributions of electrons and ions have significant effects on the dust charging process via the collections of electrons and ions in the dusty plasma.

*Additional remarks*：Most recently, the electron and ion dust charging currents are studied if the ions in the dusty plasma are the $q$-distribution[26]. The electrostatic charging of a dust particle in the electron-ion plasma[50] and the dust acoustic waves in the charge varying dusty plasma[51] are investigated if the electrons in the plasma are the $q$-distribution.

## ACKNOWLEDGMENTS

This work is supported by the National Natural Science Foundation of China under Grant No.10675088 and No.11175128.